# Dynamics of antiferromagnetic skyrmion in absence and presence of pinning defect

Z. Jin[1], T. T. Liu[1], W. H. Li[1], X. M. Zhang[1], Z. P. Hou[1], D. Y. Chen[1], Z. Fan[1], M. Zeng[1], X. B. Lu[1], X. S. Gao[1], M. H. Qin[1,3*], and J. –M. Liu[1,2]

[1]*Institute for Advanced Materials, and Guangdong Provincial Key Laboratory of Optical Information Materials and Technology, South China Academy of Advanced Optoelectronics, South China Normal University, Guangzhou 510006, China*

[2]*Laboratory of Solid State Microstructures, Nanjing University, Nanjing 210093, China*

[3]*National Center for International Research on Green Optoelectronics, South China Normal University, Guangzhou 510006, China*

**[Abstract]** A theoretical study on the dynamics of an antiferromagnetic (AFM) skyrmion is indispensable for revealing the underlying physics and understanding the numerical and experimental observations. In this work, we present a reliable theoretical treatment of the spin-current induced motion of an AFM skyrmion in the absence and presence of pinning defect. For an ideal AFM system free of defect, the skyrmion motion velocity as a function of the intrinsic parameters can be derived, based on the concept that the skyrmion profile agrees well with the 360° domain wall formula, leading to an explicit description of the skyrmion dynamics. However, for an AFM lattice containing a defect, the skyrmion can be pinned and the depinning field as a function of damping constant and pinning strength can be described by the Thiele's approach. It is revealed that the depinning behavior can be remarkably influenced by the time-dependent oscillation of the skyrmion trajectory. The present theory provides a comprehensive scenario for manipulating the dynamics of AFM skyrmion, informative for future spintronic applications based on antiferromagnets.

Email: qinmh@scnu.edu.cn

## I. Introduction

Magnetic skyrmions [1] are topologically vortex-like spin configurations, which are observable in a series of chiral magnets [2-7] and heavy metal/ferromagnetic films [8-10] with inversion symmetry breaking, and particularly in frustrated magnets [11, 12]. The interesting characters of skyrmions including nanoscale in size, topological protection character, and ultralow critical driving current [13-17], make them promising candidates for spintronic devices such as racetrack memories where the skyrmion motion becomes the core issue. Subsequently, other stimuli, such as gradient magnetic and electric fields as well as spin waves, have been proposed to effectively drive skyrmions [18-24], while the skyrmion Hall effect prohibits a precise control of the motion and even goes against their applications.

Alternatively, antiferromagnetic (AFM) skyrmions in AFM systems have been theoretically predicted [25] and have attracted extensive attention due to their particular merits. Specifically, an AFM skyrmion is comprised of two coupled spin structures with opposite topological numbers [25-27], resulting in a perfect suppression of the skyrmion Hall effect [28-33]. Thus, an AFM skyrmion can move straightly along the direction of driving stimulus without path deviation. Moreover, it has been revealed that the minimum driving current density of an AFM skyrmion is much smaller than that for a ferromagnetic (FM) skyrmion, and the velocity under the same driving force is much larger [26, 30]. In addition, strong anti-interference capability is also expected in AFM systems due to the zero stray fields [34].

For the driving stimulus, besides electric current driving, temperature gradient and magnetic anisotropy gradient [26, 31, 35] have also been theoretically uncovered to drive efficiently the motion of AFM skyrmion. For example, it is revealed that an AFM skyrmion moves towards the region of lower magnetic anisotropy under an anisotropy gradient. Furthermore, the interplay of skyrmion dynamics with defects in the system under consideration can be a core issue if the defect can pin the skyrmion motion for spintronic device operation, noting that inhomogeneity and lattice defects are inevitable in realistic materials and they may influence the magnetic dynamics by various pinning effects. Most recently, the influence of defect induced by the local variation in the magnetic anisotropy on the current-induced motion of the AFM skyrmion has been numerically investigated [36], and various behaviors of the skyrmion to these defects have been revealed. The simulated pinning and depinning of the skyrmion in response to the defects of various sizes and pinning strengths have been qualitatively explained from the energy landscape.

These important works definitely provide useful information for future AFM spintronic device design, while more comprehensive understanding of the AFM skyrmion dynamics is certainly necessary. A theoretical treatment benefits essentially to understand the physics underlying extensive numerical and experimental observations. For example, earlier simulations have demonstrated a nearly linear relation between the AFM skyrmion velocity and the Dyzaloshinskii-Moriya (DM) interaction strength [32], while the underlying physics deserves further exploration. Fortunately, earlier theoretical calculation on the FM skyrmion confirmed that the FM skyrmion profile agrees well with the $360°$ domain wall formula [37]. This equivalence allows a direct connection between the skyrmion dynamics and magnetic domain wall dynamics. Certainly, to some extent, the theoretical treatment on the FM skyrmion can be safely transferred to the case of an AFM skyrmion, considering the similarity in Hamiltonian between the FM and AFM skyrmion systems.

On the other hand, it would be even more important to understand the depinning process for an AFM skyrmion under pinning by some defect. Here, we consider the depinning process by spin current, and therefore the depinning current as a function of the system parameters such as damping constant must be well understood. In fact, in our earlier work on the depinning process of an AFM domain wall, it was predicted that the depinning field is remarkably dependent of the damping constant, attributing to the oscillation of the AFM domain wall [38]. This prediction and the equivalence mentioned just above suggests the possibility for similar dependence regarding the depinning process of an AFM skyrmion under pinning. This issue, of course, remains to be urgently clarified, no need to mention its importance in the AFM spintronics.

In this work, we focus on the spin-current-driven dynamics of an AFM skyrmion and present a reliable theoretical treatment. Then the predicted dynamic behavior will be compared with the numerical simulations based on the Landau-Lifshitz-Gilbert (LLG) equation. We start from a theoretical treatment of the dynamics of an AFM skyrmion in an ideal AFM lattice (without any pinning defect) and then the motion velocity in dependence of the intrinsic physical parameters of the system will be derived. Subsequently, we extend our treatment to the case with the presence of pinning defect in the lattice. and the depinning field of the skyrmion motion as a function of the damping constant and pinning strength of the defect will be obtained. The whole package of treatment constitutes the theory on the dynamics of AFM skyrmion in the presence of pinning defect, which is highly appreciated for potential spintronic applications.

## II. Model and methods

We consider an ultra-thin AFM film on a heavy-metal layer in the *xy* plane with two magnetic sublattices that have magnetic moments $\mathbf{m}_1$ and $\mathbf{m}_2$ respectively [39], satisfying condition $|\mathbf{m}_1| = |\mathbf{m}_2| = S$ with spin length $S$. The total magnetization $\mathbf{m}$ and the unit Néel vector $\mathbf{n}$ are defined as $\mathbf{m} = (\mathbf{m}_1 + \mathbf{m}_2)/2S$ and $\mathbf{n} = (\mathbf{m}_1 - \mathbf{m}_2)/2S$, respectively, which are used to describe the AFM skyrmion dynamics. Taking into account the exchange energy, the anisotropy energy, the interfacial DM interaction, and the constraints $|\mathbf{n}| = 1$ and $\mathbf{m} \cdot \mathbf{n} = 0$, one has the total energy of the system given by [36, 40]:

$$H = \int dV \left\{ \frac{A_0}{2} \mathbf{m}^2 + A(\nabla \mathbf{n})^2 - D[n_z \nabla \cdot \mathbf{n} - (\mathbf{n} \cdot \nabla) n_z] - K n_z^2 \right\}, \quad (1)$$

where $A_0 = 8JS^2/a^2$ is the homogeneous exchange constant with AFM interaction $J$ and lattice constant $a$, $A = JS^2$ is the inhomogeneous exchange constant, $D = D_0 S^2/a$ is the interfacial DM interaction constant with constant $D_0$ in the discrete model, $K = K_0 S^2/a^2$ is the anisotropy constant along the *z*-axis with constant $K_0$ in the discrete model [41].

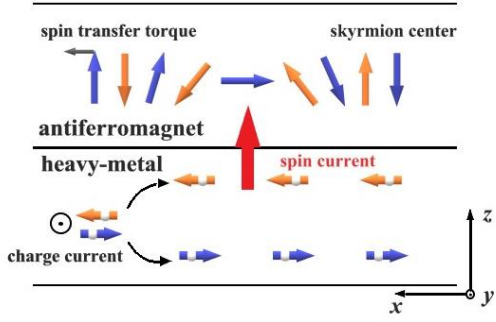

Fig. 1. Schematic illustration of an antiferromagnet-heavy metal bilayer configuration. Top layer: spin configuration along a radial direction of an AFM skyrmion. Bottom layer: a perpendicular spin current (coarse red solid arrow) is induced by a charge current (coarse black solid arrow) and then a spin transfer torque acting on AFM moment is generated.

Here, the spin-polarized current (spin current) in the perpendicular-to-plane geometry is induced by the spin Hall effect generated by the in-plane charge current in the neighboring heavy metal layer, as explicitly depicted in Fig. 1. In this case, the AFM dynamics is described by the following two coupled equations [42, 43]:

$$\dot{\mathbf{n}} = (\gamma \mathbf{f}_m - G_1 \dot{\mathbf{m}}) \times \mathbf{n} + \gamma u \mathbf{n} \times (\mathbf{m} \times \mathbf{p}), \qquad (2a)$$

$$\dot{\mathbf{m}} = (\gamma \mathbf{f}_n - G_2 \dot{\mathbf{n}}) \times \mathbf{n} + (\gamma \mathbf{f}_m - G_1 \dot{\mathbf{m}}) \times \mathbf{m} + \gamma u \mathbf{n} \times (\mathbf{n} \times \mathbf{p}), \qquad (2b)$$

where $\gamma$ is the gyromagnetic ratio, $\mathbf{f}_m = -\delta H/\delta \mathbf{m}$ and $\mathbf{f}_n = -\delta H/\delta \mathbf{n}$ are the effective fields, $G_1$ and $G_2$ are the phenomenological Gilbert damping parameters, $\mathbf{p}$ denotes the unit vector along the electron polarization direction ($x$-axis, $\mathbf{e}_x$), $u = j\mu_B\theta_{SH}/(\gamma e\mu_s t_z)$ is the effective field related to the damping-like spin torque with the Bohr magneton $\mu_B$, the effective spin-Hall angle $\theta_{SH}$, the driving current density $j$, the electron charge $e$, the saturation moment $\mu_s$, and the film thickness $t_z$. Subsequently, the dynamics of the AFM skyrmion can be analytically calculated by solving the two equations.

Furthermore, the velocity and the depinning field of the AFM skyrmion motion are also estimated using the LLG simulations of the discrete model, in order to check the validity of the theoretical treatment. The simulation details and parameter choice are presented in Appendix A.

## III. Results and discussion

### A. Velocity for AFM skyrmion motion

In the earlier work [32], it was numerically revealed that the velocity of an AFM skyrmion is proportional to the DM interaction for a fixed current density. Here, we can analytically derive the relationship between the two parameters which is nearly linear, in order to uncover the physical pictures behind the earlier simulations.

By substituting Eq. (2a) into Eq. (2b) and neglecting the nonlinear terms, one obtains the dynamic equation expressed by the Néel vector **n**:

$$(1+G_1G_2)\dot{\mathbf{n}} = -A_0\gamma \mathbf{m}\times\mathbf{n} + \gamma G_1\mathbf{f_n} + G_1\gamma u(\mathbf{p}\times\mathbf{n}), \qquad (3)$$

noting that the effective fields are time independent. Then, by taking a derivative with respect to time and safely neglecting small terms, we obtain:

$$\frac{\ddot{\mathbf{n}}}{\gamma^*} = A_0(\gamma \mathbf{f_n} - G_2\dot{\mathbf{n}} + \gamma u(\mathbf{p}\times\mathbf{n})), \qquad (4)$$

where $\gamma^* \equiv \gamma/(1+G_1G_2) \approx \gamma$.

Subsequently, we investigate the dynamics using the collective coordinate approach [44, 45], and obtain the velocity $v$ for an AFM skyrmion moving along the current direction without the skyrmion Hall effect,

$$v = \frac{\gamma I_{xy} u}{G_2 D_{xx}}, \qquad (5)$$

where $I_{xy}$ and $D_{xx}$ are the driving force tensor and dissipative tensor, respectively. For the convenience of mathematical treatment, the Néel vector at point $r$ is described by $\mathbf{n} = (\sin\theta\cos\varphi, \sin\theta\sin\varphi, \cos\theta)$ with the polar angle $\theta(r)$ and azimuthal angle $\varphi = \Phi$ (the azimuthal angle of $r$), considering a skyrmion centered at $r = 0$. Thus, the tensors $I_{xy}$ and $D_{xx}$, respectively, read [46]:

$$I_{xy} = \int(\frac{\partial n_z}{\partial x}n_x - \frac{\partial n_x}{\partial x}n_z)dxdy = \pi\int r(\frac{\partial \theta}{\partial r} + \frac{\sin 2\theta}{2r})dr, \qquad (6)$$

And

$$D_{xx} = \int [(\frac{\partial n_x}{\partial x})^2 + (\frac{\partial n_y}{\partial x})^2 + (\frac{\partial n_z}{\partial x})^2]dxdy = \pi \int r[(\frac{\partial \theta}{\partial r})^2 + \sin^2\theta/r^2]dr, \tag{7}$$

It is seen that the magnetization profile along the radial direction of a FM or AFM skyrmion is equivalent to a 360° domain wall. Thus, the 360° domain wall could be used to describe the skyrmion profile, as confirmed in earlier work on the FM skyrmion size [37]. As a matter of fact, this conceptual scenario can be safely transferred to the study of the AFM skyrmion dynamics. By this equivalence, the skyrmion profile reads in this case:

$$\theta(r) = 2\arctan[\frac{\sinh(R_s/w)}{\sinh(r/w)}], \tag{8}$$

where the wall width $w = \pi D/4K$, and the skyrmion size $R_s = \pi D[A/(16AK^2 - \pi^2 D^2 K)]^{1/2}$ with the constraint $16AK > \pi^2 D^2$. One obtains by substituting Eq. (8) into Eq. (6) and (7):

$$I_{xy} = 2\pi w \int_0^\infty [-\frac{\sinh t_2 \cosh t_1}{\sinh^2 t_2 + \sinh^2 t_1}t_1 - \frac{\sinh t_2 \sinh t_1(\sinh^2 t_2 - \sinh^2 t_1)}{(\sinh^2 t_2 + \sinh^2 t_1)^2}]dt_1, \tag{9}$$

and

$$D_{xx} = 2\pi \int_0^\infty \{\frac{2\sinh^2(t_2)\cosh^2(t_1)}{[\sinh^2(t_2) + \sinh^2(t_1)]^2}t_1 + \frac{2\sinh^2(t_2)\cosh^2(t_1)}{[\sinh^2(t_2) + \sinh^2(t_1)]^2 t_1}\}dt_1, \tag{10}$$

where $t_1 = R_s/w$, $t_2 = r/w$. For $r \gg w$, one has $\sinh(r) \approx \cosh(r) \approx e^r$. Thus, we obtain $I_{xy}$ and $D_{xx}$:

$$\begin{aligned} I_{xy} &= \pi^2 R_s \\ D_{xx} &= 2\pi(\frac{R_s}{w} + \frac{w}{R_s}) \end{aligned}, \tag{11}$$

Substituting Eq. (11) into Eq. (5) yields

$$v = \frac{2DA\pi^2 \gamma}{(32AK - \pi^2 D^2)G_2}u, \tag{12}$$

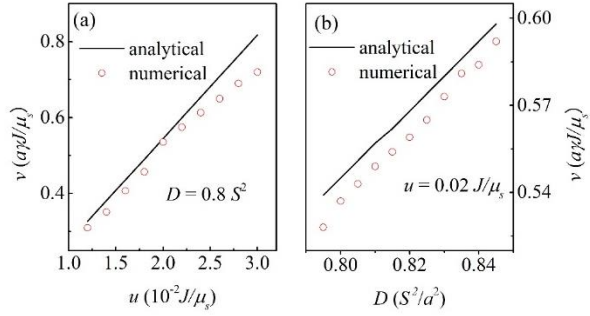

Fig. 2. The simulated (empty circles) and analytically calculated (solid line) velocity $v$ as functions of $u$ for $D = 0.8\ S^2$ (a), and as a function of $D$ for $u = 0.02\ J/\mu_s$ (b).

In order to check the validity of theory, a comparison between the analytical calculations and numerical simulations is indispensable. Fig. 2(a) gives the calculated and LLG simulated $v$ as a function of $u$ for $D = 0.8 S^2/a^2$. The spin transfer torque is enhanced with the increase in $u$, and drives the skyrmion to move fast. The theory prediction coincides well with the simulated ones especially for the cases of not large $u$, confirming the validity of the theory. It is noted that the AFM skyrmion could be deformed during its motion if the driving current in the simulation is very large, resulting in the discrepancy between the calculations and simulations in the large $u$ range. Similar to the AFM domain wall, the speed of AFM skyrmion is also limited [47], attributing to the Lorentz contraction. Furthermore, the DM interaction determines the skyrmion size and in turn affects the skyrmion velocity. The velocity increases monotonically with the DM interaction, as shown in Fig. 2(b) which presents the calculated and simulated $v$ as a function of $D$ for $u = 0.02\ J/\mu_s$. The results are in well consistent with each other (the discrepancy is less than 5%), further confirming the validity of the theory. Moreover, for $32AK \gg \pi^2 D^2$, a nearly linear relation between $v$ and $D$ is obtained, quantitatively explaining the earlier simulations [32].

So far, the dependence of the AFM skyrmion velocity on these internal and external parameters is clarified, allowing one to estimate the velocity easily and to understand the physics clearly. As a matter of fact, the dynamics of AFM skyrmion has attracted attention for many years, but a detailed formulation for its velocity remained ambiguous before the present work.

## B. Pinning and depinning of AFM skyrmion

Subsequently, we investigate the AFM skyrmion dynamics in a system with pinning defect, in particular the pinning and depinning behaviors. Without loss of generality, the defect is introduced by a local variation in the magnetic anisotropy [36, 48, 49], expressed by $K^* = K\{1 - \lambda \exp[-(r - r_d)^2/R_d^2]\}$ [26], where $\lambda$ denotes the defect pinning strength, $r_d$ is the position of the defect center, and $R_d$ is the defect size.

For our calculation, parameters $K = 0.8S^2/a^2$ and $D = 0.78S^2/a$ are selected to generate a skyrmion with small size to suppress the defect induced skyrmion distortion. Thus, for $R_s < R_d$, one may reasonably assume the defect potential to be a parabolic one [38, 50]:

$$V(r) = \begin{cases} \frac{1}{2}\lambda_0 |r - r_d|^2 & (|r - r_d| < R_p) \\ \frac{1}{2}\lambda_0 R_p^2 & (|r - r_d| \geq R_p) \end{cases}, \tag{13}$$

where $\lambda_0 = c_\lambda \lambda$ is the pinning strength prefactor related to the defect, and $R_p$ is the radius of the potential well.

For simplicity, the skyrmion is assumed to be initially at the defect center $r_d$. It is noted that the skyrmion will be captured by the defect in the low current region. By applying the Thiele approach and considering Eq. (13), the equation of motion for the skyrmion position $q$ is obtained:

$$\ddot{q} + 2\varepsilon\omega\dot{q} + \omega^2 q + C = 0, \tag{14}$$

where $2\varepsilon = \gamma A_0 G_2 / \omega$, $\omega^2 = \gamma c_\lambda \lambda / D_{xx}$, and $C = -\gamma^2 A_0 I_{xy} u/D_{xx}$. Eq. (14) clearly describes the damping oscillation of the AFM skyrmion. Specifically, for $G_2 < 2\omega/\gamma A_0$, we have the solution representing an underdamped oscillation:

$$q(\tau) = e^{-\varepsilon\omega\tau}(C_1 \cos\omega_p\tau + C_2 \sin\omega_p\tau) - \frac{C}{\omega^2}, \tag{15}$$

where $C_1$ and $C_2$ are parameters to be determined by initial conditions, $\omega_p = (\omega^2 - (\gamma A_0 G_2)^2/4)^{1/2}$ is the oscillating angular frequency, and $\omega_p \approx \omega$ due to $\omega^2 \gg (\gamma A_0 G_2)^2/4$. It is noted that $G_1$ is much smaller than $G_2$ [42], which has been safely neglected in the derivation of Eq. (15).

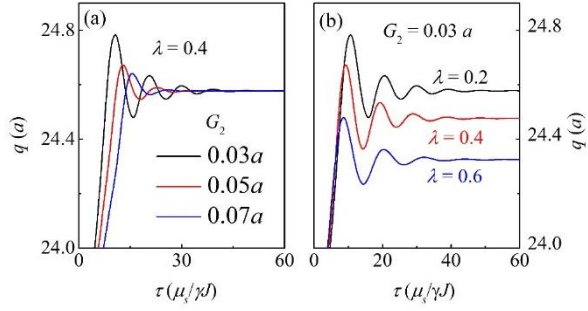

Fig. 3. The skyrmion position as a function of time (a) for various $G_2$ for $\lambda = 0.4$, and (b) for various $\lambda$ for $G_2 = 0.03\ a$ under $u = 0.012\ J/\mu_s$.

As a matter of fact, the underdamped oscillation has been confirmed by the LLG simulations, as shown in Fig. 3(a) that presents the simulated $q(\tau)$ curves for various $G_2$. For a fixed $G_2$, the skyrmion oscillates around its equilibrium position with an attenuating amplitude. It is noted that at the equilibrium position, the driving torque is well cancelled out by the retarding torque caused by the defect, and the position hardly depends on $G_2$. Interestingly, the oscillation magnitude is highly related to the damping constant, which significantly affects the pinning and depinning of the skyrmion. Actually, one may define the maximum displacement of the skyrmion from its initial position as $|\Delta q|_{max}$, which is approximately given by:

$$|\Delta q|_{max} = e^{-\gamma A_0 G_2 \arctan(C_2/C_1)/2\omega_p}\sqrt{C_1^2 + C_2^2} - \frac{C}{\omega^2}, \qquad (16)$$

The simulated $|\Delta q|_{max}$ for various $G_2$ are summarized in Fig. 4(a), in which the excellent fitting of the simulated data based on Eq. 16 confirms the validity of the theory.

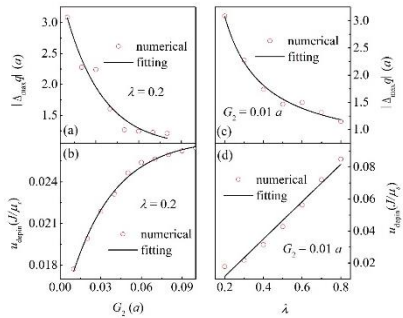

Fig. 4. Numerical (empty circles) and analytical (solid line) calculated $|\Delta q|_{max}$ ((a) and (c)), and $u_{depin}$ ((b) and (d)) as functions of $G_2$ for $\lambda = 0.2$ ((a) and (b)) and as functions of $\lambda$ for $G_2 = 0.01\ a$ ((c) and (d)).

When a large current is applied, the AFM skyrmion will be depinned from the defect, benefiting from the strong driving torque. The critical depinning field could be theoretically estimated based on the condition that the maximum displacement of the skyrmion equals to the radius of the potential well, $|\Delta q|_{\max} = R_p$. Substituting the condition into Eq. 16, one obtains the critical depinning field:

$$u_{\text{depin}} = (R_p - e^{-\gamma A_0 G_2 \arctan(C_2/C_1)/2\omega_p} \sqrt{C_1^2 + C_2^2}) \omega^2 D_{xx} / \gamma^2 A_0 I_{xy}, \tag{17}$$

The simulated $u_{\text{depin}}$ as a function of $G_2$ are plotted in Fig. 4(b), which reveals a significant dependence of $u_{\text{depin}}$ on $G_2$ in its small region. Specifically, $u_{\text{depin}}$ gradually increases with $G_2$ until the large $G_2$ region where $u_{\text{depin}}$ becomes saturated, similar to the depinning of the AFM domain wall in notched nanostructures [38]. Importantly, the simulated results are well in consistent with Eq. 17, revealing that the oscillation of the AFM skyrmion plays an important role in its depinning, regardless of the defect type.

Unlike the damping constant, the defect strength $\lambda$ also affects the equilibrium position of the skyrmion, as revealed in Eq. 15. For example, for a small $\lambda$, the skyrmion moves to an equilibrium position far away from its initial position to make the torque from the defect cancels out the driving torque, as shown in Fig. 3(b) where presents the simulated $q(\tau)$ curves for various $\lambda$. In this case, the oscillation magnitude hardly depends on $\lambda$ for small damping constant, allowing one to fit the simulated $|\Delta q|_{\max}$ by $|\Delta q|_{\max} = C^* - CD_{xx}/c_\lambda\lambda\gamma$ where constant $C^*$ represents the first term on the right side of Eq. 16. The updated equation of $|\Delta q|_{\max}$ coincides well with the simulated results, as shown in Fig. 4(c) where presents the LLG-simulated $|\Delta q|_{\max}$ as a function of $\lambda$ for $G_2 = 0.01a$. Similarly, the equation of $u_{\text{depin}}$ could be updated to $u_{\text{depin}} = (R_p - C^*) c_\lambda\lambda/I_{xy}\gamma A_0$, revealing a linear relation between $u_{\text{depin}}$ and $\lambda$. This relation has been verified by the fitting of the simulated $u_{\text{depin}}$ as a function of $\lambda$, as shown in Fig. 4(d).

## C. Brief discussion

In part III. A, the dependence of the skyrmion velocity on the internal parameters including the exchange interaction and DM interaction is predicted, based on the fact that the AFM skyrmion profiles agree with the 360° domain wall formula. In part III. B, the critical depinning field of the AFM skyrmion is discussed, based on the Thiele's theory, and its dependence on the damping constant and pinning strength is derived. The theory predictions are well consistent with the simulation results based on the LLG equation, confirming the validity of the theory.

On one hand, the deformation of the AFM skyrmion in shape could be induced during the motion or by the interaction with the defect. For simplicity, the deformation [32] is completely ignored in our derivation of the dynamics, resulting in the small deviation between the simulations and analytical calculations. It is noted that the skyrmion size significantly depends on the exchange and DM interactions which could be also changed in the defect region. Their effects on the depinning of the skyrmion have been numerically investigated in earlier numerical simulations [36], which is beyond the scope of this theory. Moreover, the defect potential could be complex when the defect size is smaller than the skyrmion size. Thus, the present theory could work well for the skyrmion with a small and stable size [51].

On the other hand, the spin current, as an example, is used to drive the motion of the AFM skyrmion in this work. However, the theory could be easily transferred to other proposed schemes for driving the AFM skyrmion motion through re-calculating the effective field. More interestingly, a parabolic potential is also expected for other types of defects such as notches [38, 52], and the derived formula for calculating the depinning field could also work in notched nanostructures, which deserves to be checked in future simulations and/or experiments [53].

Thus, the present theory not only strengthens the earlier conclusions, but more importantly helps one to understand the physical picture behind the numerical simulations, providing useful information for future application design.

**IV. Conclusion**

In conclusion, we have studied theoretically the spin-current-induced dynamics of the AFM skyrmion, and have confirmed the validity of the theory through a detailed comparison between the theory and numerical simulations based on the LLG equation. The dependence of the skyrmion velocity on the intrinsic physical parameters are uncovered, allowing a quantitatively and clearly understanding of the dynamics. Moreover, the depinning field of the AFM skyrmion depending on the damping constant and the defect's pinning strength is derived theoretically, demonstrating the effect of the time-dependent oscillation of the skyrmion on the depinning. Thus, the present work helps one to understand clearly the velocity and defect-depinning of the AFM skyrmion, benefiting future experiments designs and AFM spintronics applications.

## Acknowledgment

We sincerely appreciate the insightful discussions with Xichao Zhang, Huaiyang Yuan, and Laichuan Shen. The work is supported by the Natural Science Foundation of China (No. 51971096, 51721001), and the Natural Science Foundation of Guangdong Province (Grant No. 2019A1515011028), and the Science and Technology Planning Project of Guangzhou in China (Grant No. 201904010019), and Special Funds for the Cultivation of Guangdong College Students Scientific and Technological Innovation (Grant No. pdjh2020a0148).

**Appendix A: Numerical simulations of the atomistic spin model**

In order to check the validity of the theory, we also perform the numerical simulations of the discrete model. Here, the two dimensional Hamiltonian of the atomistic spin model is given by

$$H = J\sum_{\langle i,j \rangle} \mathbf{S}_i \cdot \mathbf{S}_j - D_0 \sum_i (\mathbf{S}_i \times \mathbf{S}_{i+x} \cdot \hat{\mathbf{y}} - \mathbf{S}_i \times \mathbf{S}_{i+y} \cdot \hat{\mathbf{x}}) - K_0 (S_i^z)^2, \tag{A1}$$

where the first term is the exchange interaction with $J = 1$ between the nearest neighbor spins, the second term is the DM interaction, and the last term is the anisotropy energy. The dynamics of the AFM skyrmion driven by the current is investigated by solving the LLG equation,

$$\frac{\partial \mathbf{S}_i}{\partial \tau} = -\gamma \mathbf{S}_i \times \mathbf{H}_i + \alpha \mathbf{S}_i \times \frac{\partial \mathbf{S}_i}{\partial \tau} + \gamma u \mathbf{S}_i \times (\mathbf{S}_i \times \mathbf{p}), \tag{A2}$$

where $\alpha$ is the damping constant, $\mathbf{H}_i = -\mu_s^{-1} \partial H / \partial \mathbf{S}_i$ is the effective field. Without loss of generality, $u = 0.02\ J/\mu_s$, $K_0 = 0.8\ J$ and $\alpha = 0.03$ are selected. Generally, we use the fourth-order Runge-Kutta method to solve the LLG equation on a $50 \times 50$ square lattice. The position of the skyrmion $q$ is estimated by

$$q = \frac{\int [x \mathbf{n} \cdot (\partial_x \mathbf{n} \times \partial_y \mathbf{n})] dx dy}{\int [\mathbf{n} \cdot (\partial_x \mathbf{n} \times \partial_y \mathbf{n})] dx dy}, \tag{A3}$$

Then, the velocity is numerically calculated by $v = dq/d\tau$ with time $\tau$.